\documentclass[fleqn,twoside]{article}
\usepackage{espcrc2}
\usepackage{graphicx}
\usepackage[normalem]{ulem}
\usepackage{url}
\urldef{\het}\url{www.het.brown.edu}
\newcommand\email{\begingroup \urlstyle{tt}\Url}
\urldef{\danieldf}{\email}{danieldf@het.brown.edu}
\usepackage[dvips,usenames]{color}
\usepackage[american]{babel}
\usepackage[latin1]{inputenc}
\usepackage[T1]{fontenc}
\usepackage{xspace}
\usepackage{latexsym}
\usepackage{amsmath,amsfonts,amstext,amssymb,amsbsy,amsopn,amsthm,eucal}
\usepackage{yfonts}[1998/10/03]
\usepackage{dsfont}
\usepackage{wasysym}
\usepackage{wrapfig}
\usepackage{algorithm}
\usepackage{algorithmic}
\DeclareMathOperator{\Ai}{Ai}

\newcommand{\varsqrt}[1]{\ensuremath{\sqrt{#1{\,}}}\xspace}
\newcommand{\conv}[2]{\ensuremath{{#1}\ast{#2}}\xspace} 

\newcommand{\dslash}{\not{\hbox{\kern-2pt $\partial$}}}
\newcommand{\pslash}{\not{\hbox{\kern-2.3pt $p$}}}
 \newtoks\nslashfraction
 \nslashfraction={.13}
 \newcommand{\nslash}[1]{\setbox0\hbox{$ #1 $}
   \setbox0\hbox to \the\nslashfraction\wd0{\hss \box0}/\box0 }
\newcommand{\plpl}{\raise-2pt\hbox{$\raise3pt\hbox{$_+$}\hskip-6.67pt\raise0.0pt
  \hbox{$^+$}\hskip 0.01pt$}}
\newcommand{\mimi}{\raise-2pt\hbox{$\raise3pt\hbox{$_-$}\hskip-6.67pt\raise0.0pt
  \hbox{$^-$}\hskip 0.01pt$}}
\newcommand{\bo}{\raise-1mm\hbox{\Large$\Box$}}              
\newcommand{\pa}{\partial}                                       
\newcommand{\trans}[1]{{#1}^{\ensuremath{\mathsf{T}}}}           
\newcommand{\nTH}{{\raise.2ex\hbox{$\displaystyle \bigodot$}\mskip-4.7mu \llap H \;}}
\newcommand{\face}{{\raise.2ex\hbox{$\displaystyle \bigodot$}\mskip-2.2mu \llap {$\ddot
        \smile$}}}                                      

\newcommand{\ev}[1]{\left\langle #1\right\rangle}        
\newcommand{\abs}[1]{\left| #1\right|}                    

\newcommand{\leftrightarrowfill}{$\mathsurround=0pt \mathord\leftarrow \mkern-6mu
        \cleaders\hbox{$\mkern-2mu \mathord- \mkern-2mu$}\hfill
        \mkern-6mu \mathord\rightarrow$}
\newcommand{\dvec}[1]{\vbox{\ialign{##\crcr
        \leftrightarrowfill\crcr\noalign{\kern-1pt\nointerlineskip}
        $\hfil\displaystyle{#1}\hfil$\crcr}}}           

\newcommand{\sfrac}[2]{{\vphantom1\smash{\lower.5ex\hbox{\small$#1$}}\over
        \vphantom1\smash{\raise.4ex\hbox{\small$#2$}}}} 
\newcommand{\bfrac}[2]{{\vphantom1\smash{\lower.5ex\hbox{$#1$}}\over
        \vphantom1\smash{\raise.3ex\hbox{$#2$}}}}       
\newcommand{\afrac}[2]{{\vphantom1\smash{\lower.5ex\hbox{$#1$}}\over#2}}    
\newskip\humongous \humongous=0pt plus 1000pt minus 1000pt

\newif\ifdtup

\newfont{\go}{ygoth.tfm scaled 1200}                   
\newfont{\biggo}{ygoth.tfm scaled 3583}                
\newfont{\rope}{cmsy10 scaled 1200}                    
\newfont{\fib}{cmfi10 scaled 1200}
\newfont{\bigfib}{cmfi10 scaled 3583}
\newfont{\funny}{cmff10 scaled 1200}
\newfont{\bigfunny}{cmff10 scaled 3583}
\newfont{\pbk}{pbkd.tfm scaled 1200}
\newfont{\rsfs}{rsfs10.tfm scaled 1200}
\newfont{\bigrsfs}{rsfs10.tfm scaled 2000}
\newfont{\testea}{cmfrak.tfm}
\newfont{\testeb}{dcfrak.tfm}
\newfont{\testec}{schwell.tfm}
\newfont{\tested}{yfrak.tfm}
\newfont{\testef}{yswab.tfm}
\newfont{\bigtestef}{yswab.tfm scaled 3583}
\newfont{\testeg}{yinit.tfm}
\newfont{\testeh}{yinitdd.tfm}
\newfont{\testei}{suet14.tfm}
\newfont{\testej}{pzdr.tfm}
\newfont{\testek}{pzcmi.tfm}
\newfont{\testem}{ccr10.tfm}
\newfont{\testen}{eurm10.tfm}
\newfont{\testeq}{euex10.tfm}
\newfont{\testeo}{wncyr10.tfm}
\newfont{\testep}{msam10.tfm}
\catcode`@=11
\newcommand{\un}[1]{\relax\ifmmode\@@underline#1\else
        $\@@underline{\hbox{#1}}$\relax\fi}
\catcode`@=12

\title{Mollified Monte Carlo \hfill\small \texttt{hep-lat/0209053}, BROWN-HET-1317}
\author{D. D. Ferrante\thanks{\danieldf} \address[HET]{Physics Department, Brown University, \\
    Providence, RI. 02912. POBOX 1843.}, J. Doll \address[CHEM]{Chemistry Department,
    Brown University, \\ Providence, RI. 02912.}, G. S. Guralnik \addressmark[HET],
  D. Sabo\addressmark[CHEM].}
\begin{document}
\begin{abstract}
  Using a common technique \cite{pde,fismath} for approximating distributions
  [generalized functions], we are able to use standard Monte Carlo methods to
  compute QFT quantities in Minkowski spacetime, under phase transitions, or when
  dealing with coalescing stationary points.
\end{abstract}
%
%
\maketitle
\section{Theory}
We use the mollification (\emph{approximate identity}) technique to rewrite the partition
function in such a way that not only its computation, but the computation of its
derivatives (Green's functions), become well defined using Monte Carlo methods, even for
integrands that normally do not allow this approach.
\subsection{Mollification}
A mollifier, $\eta$, is a positive, $C^{\infty}[I]$ function, $(I \in \mathbb{R})$, with
$\text{supp}(\eta) = B(0,l)$, where $l = \text{length}(I)$ and $B(0,l)$ is the ball centered
at $0$ with radius $l$ and $\int_I \eta = 1$. The sequence of functions, $\eta_{\epsilon}
= \epsilon^{-n}\, \eta(x/\epsilon)$ is an \emph{approximate identity}. A
\emph{mollification} of a locally integrable function $f:\, U\rightarrow \mathbb{R}$,
$(U\in \mathbb{R}^n, \text{ open})$, is given by,
\begin{flalign}
  \nonumber
  U_{\epsilon} &= \{x \in U\; \brokenvert\; \text{dist}(x,\pa U) > \epsilon\} \\
  \label{eq:moll}
  f_{\epsilon} &= \conv{\eta_{\epsilon}}{f} \; , \quad f_{\epsilon}\in U_{\epsilon} \\
  \label{eq:moll-conv}
  f_{\epsilon}(x) &= \int_U\, \eta_{\epsilon}(x-y)\, f(y)\, dy \\
  \nonumber
  &= \int_{B(0,\epsilon)}\, \eta_{\epsilon}(y)\, f(x-y)\, dy \; .
\end{flalign}
\subsection{Properties}
The properties of mollifiers, $(f_{\epsilon})$, are:
\begin{enumerate}
  \item $f_{\epsilon}\in C^{\infty}(U_{\epsilon})$;
  \item $f_{\epsilon} \rightarrow f$, almost everywhere, as $\epsilon\rightarrow 0$;
  \item If $f\in C(U)$, then $f_{\epsilon} \rightarrow f$
    \emph{uniformly} on compact subsets of $U$; \emph{\&}
  \item If $1 \leqslant p < \infty$ and $f\in L_{\text{loc}}^p(U)$, then $f_{\epsilon}
    \rightarrow f$ in $L_{\text{loc}}^p(U)$.
\end{enumerate}
It is not difficult to see that this approximation, ($f_{\epsilon}$, the
\emph{mollified} version of $f$), can smooth oscillating functions. $f_{\epsilon}$ is
differentiable and, not only does it converge to $f$ when $\epsilon\rightarrow 0$
but, depending on where this convergence takes place, it can be either uniform or in
$L_{\text{loc}}^p$.
%
%
%
\subsection{Implementation of the Method}
The object we calculate will not be the usual
partition function, but its mollified version, $\mathcal{Z}_{\epsilon}$,
\begin{flalign}
  \label{eq:pf}
  \mathcal{Z}_{\epsilon}[J] &= \frac{\displaystyle\varint\, \ev{e^{i\, S(\mathbf{\phi}) -
          i\, J\cdot\phi}}_{\epsilon}\, [d\phi]}{\displaystyle\varint\, \ev{e^{i\,
          S(\mathbf{\phi})}}_{\epsilon}\, [d\phi]} \;\; , \\
  \intertext{where,}
  \nonumber
  \ev{f}_{\epsilon} &\equiv f_{\epsilon} \equiv \conv{\eta_{\epsilon}}{f} \; .
\end{flalign}
From equation \eqref{eq:moll-conv} it follows that, upon choosing $U$ appropriately, the
boundary conditions\footnote{Note that, a choice of boundary conditions [Schwinger-Dyson
  equation] is equivalent to choosing the measure in the path-integral.} for the theory
(defined here by $S$) are set\cite{gerry}.

The method is applied in two steps:
\begin{enumerate}
\item \uline{Saddle-point expansion}: For simplicity, the mollifier is chosen to be a
  Gaussian\footnote{Note that, bold symbols and bold letters, denote vector/matrix
    quantities.},
  \begin{equation}
    \label{eq:gaussian-moll}
    \eta_{\epsilon}(\boldsymbol{\varphi}) = \frac{\exp\Bigl\{-\frac{1}{2}\,
          \trans{\boldsymbol{\varphi}}\cdot (\boldsymbol\epsilon^2)^{-1}\cdot
          \boldsymbol{\varphi} \Bigr\}}{\varsqrt{(2\pi)^n\, \det(\boldsymbol\epsilon^2)}}
          \; ,
  \end{equation}
  where $\boldsymbol\epsilon^2$ is the covariance matrix that defines the Gaussian
  distribution. Thus, a typical path-integral will look like,
  \begin{flalign*}
    \mathcal{I}_{\epsilon} &= \varint \ev{f(\boldsymbol{\varphi})\,
      e^{i\,S(\boldsymbol{\varphi}) - i\, J\cdot \boldsymbol{\varphi}}}_{\epsilon} \,
      [d\varphi] \\
    &= f(\boldsymbol{\varphi}_0)\, e^{i\,S(\boldsymbol{\varphi}_0) - i\,
      J\cdot\boldsymbol{\varphi}_0}\, \times \\
    &\hspace*{-0.3cm}{\footnotesize \times \varint\, \frac{\exp\Bigl\{-\frac{1}{2}\,
      \trans{\mathds{B}}\cdot (\mathds{1 + \trans{\boldsymbol\epsilon}\cdot
      \mathds{S}''\cdot \boldsymbol\epsilon})^{-1}\cdot
      \mathds{B}\Bigr\}}{\varsqrt{\det(\mathds{1} + \trans{\boldsymbol\epsilon}\cdot
      \mathds{S}''\cdot \boldsymbol\epsilon)}} \, [d\varphi] } + \\
    &+ \text{corrections} \; ,
  \end{flalign*}
  where $\mathds{B} = \boldsymbol{\epsilon\cdot \nabla} S(\boldsymbol{\varphi})$ and
    $[\mathds{S}'']_{ij} = \partial^2 S(\boldsymbol{\varphi})/\partial \varphi_i\,
    \partial \varphi_j$ (Hessian matrix for $S$).

\item \uline{Importance sampling}: The Monte Carlo simulation is handled as
  follows: Choose an \emph{Importance Function}, $(W(\varphi))$, and change the measure of
  integration accordingly,
  \begin{equation}
    \label{eq:ismc}
    \mathcal{Z}_{\epsilon}[J] = \frac{\displaystyle\varint\, \biggl[\frac{\ev{e^{i\,
      S(\boldsymbol{\varphi}) - i\, J \cdot
      \boldsymbol{\varphi}}}_{\epsilon}}{W(\boldsymbol{\varphi})}\biggr]\, [d
      W(\boldsymbol{\varphi})]}{\displaystyle\varint\, \biggl[\frac{\ev{e^{i\, 
      S(\boldsymbol{\varphi})}}_{\epsilon}}{W(\boldsymbol{\varphi})}\biggr]\, [d
      W(\boldsymbol{\varphi})]} \;\; .
  \end{equation}
  A reasonable choice for $W$ is given by \cite{doll} $W(\boldsymbol\varphi) =
  W_{\epsilon}(\boldsymbol{\varphi}) = \abs{\ev{e^{i\,
  S(\boldsymbol{\varphi})}}_{\epsilon}}$. Upon a saddle-point expansion, we find,
  {\footnotesize
  \begin{equation}
    \label{eq:impsampchoice}
    W_{\epsilon}(\boldsymbol{\varphi}) = \abs{ \frac{ \exp \Bigl\{ i\,
      S(\boldsymbol{\varphi}_0) - \frac{1}{2}\, \trans{\mathds{B}}\cdot (\mathds{1 +
      \trans{\boldsymbol\epsilon}\cdot \mathds{S}''\cdot \boldsymbol\epsilon})^{-1}\cdot
      \mathds{B}\Bigr\}}{\varsqrt{\det(\mathds{1} + \trans{\boldsymbol\epsilon}\cdot
      \mathds{S}''\cdot \boldsymbol\epsilon)}} } \; .
  \end{equation}
  }
\end{enumerate}
\section{Applications in Physics}
Two basic examples will be shown here. Both of them can be regarded as
0-dimensional [spacetime] quantum field theories.
\subsection{Airy Action}
The action is given by: $S(x) = i\, x^3/3 + i\, t\, x$. Note that,
$\int_{-\infty}^{\infty} \exp\{S(x)\}\, dx = \Ai(t)$, thus, the partition function and its
mollification are,
\begin{flalign}
  \label{aipf}
  \mathcal{Z}[t] &= \frac{\displaystyle\varint_{-\infty}^{\infty}\, \exp\biggl\{i\,
      \frac{x^3}{3} + i\, t\, x\biggr\}\, dx}{\displaystyle\varint_{-\infty}^{\infty}\,
      \exp\biggl\{i\, \frac{x^3}{3}\biggr\}\, dx} \equiv \frac{\Ai(t)}{\Ai(0)} \; , \\
  \label{aimpf}
  \mathcal{Z}_{\epsilon}[t] &= \frac{\displaystyle\varint_{-\infty}^{\infty}\,
      \ev{\exp\biggl\{i\, \frac{x^3}{3} + i\, t\, x\biggr\}}_{\epsilon}\,
      dx}{\displaystyle\varint_{-\infty}^{\infty}\, \ev{\exp\biggl\{i\,
      \frac{x^3}{3}\biggr\}}_{\epsilon}\, dx} \; .
\end{flalign}
This is a particular useful example for two reasons:
\begin{enumerate}
\item We can easily compare the approximated results with the exact ones,
\item Boundary Conditions: Assuming integration on the real axis, the Euclidian version of
  this theory does not exist, therefore working in Minkowski space is necessary.
\end{enumerate}
It is easy to see that this is a highly oscillatory integrand (what makes the use of Monte
Carlo techniques impractible). However, once its mollification is calculated, the task
becomes simpler (as shown below).
\begin{center}
  \hspace{\fill}
  \scalebox{0.2}{\includegraphics{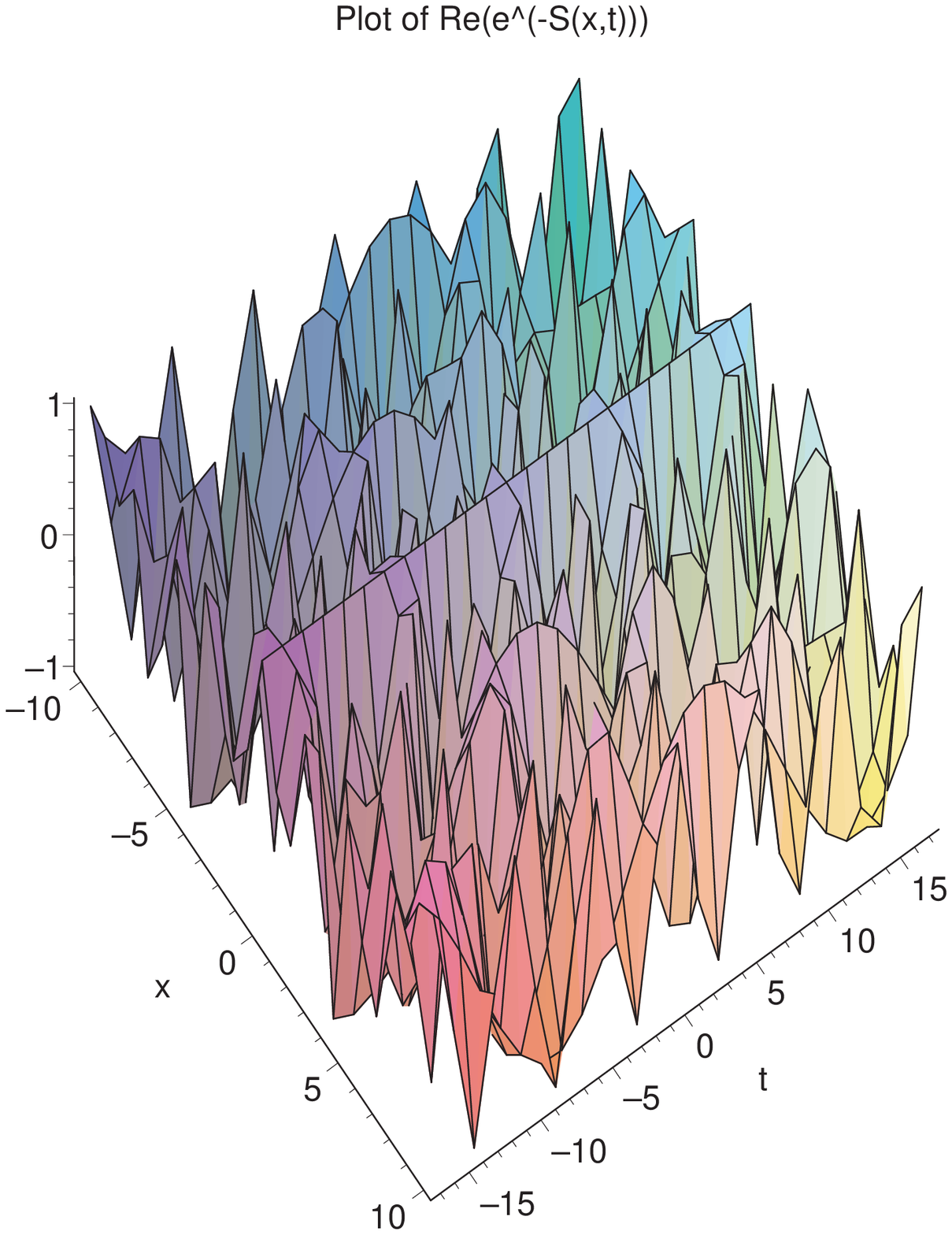}}
  \hspace{\fill}
  \scalebox{0.2}{\includegraphics{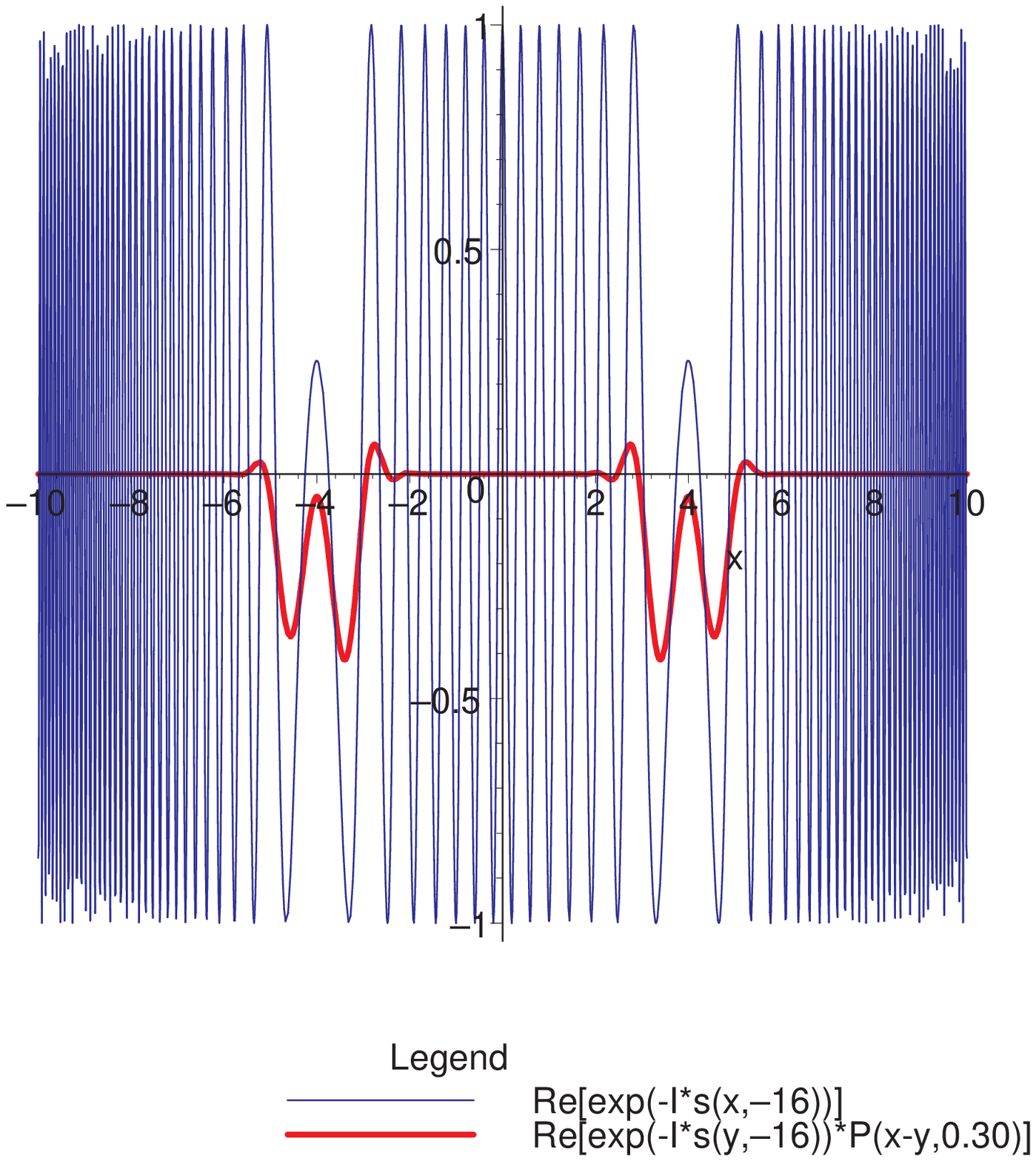}}
  \hspace{\fill}
\end{center}
The importance function, $(W(x))$, is also shown for several values of the parameter
$\epsilon$:
\begin{center}
  \scalebox{0.3}{\includegraphics{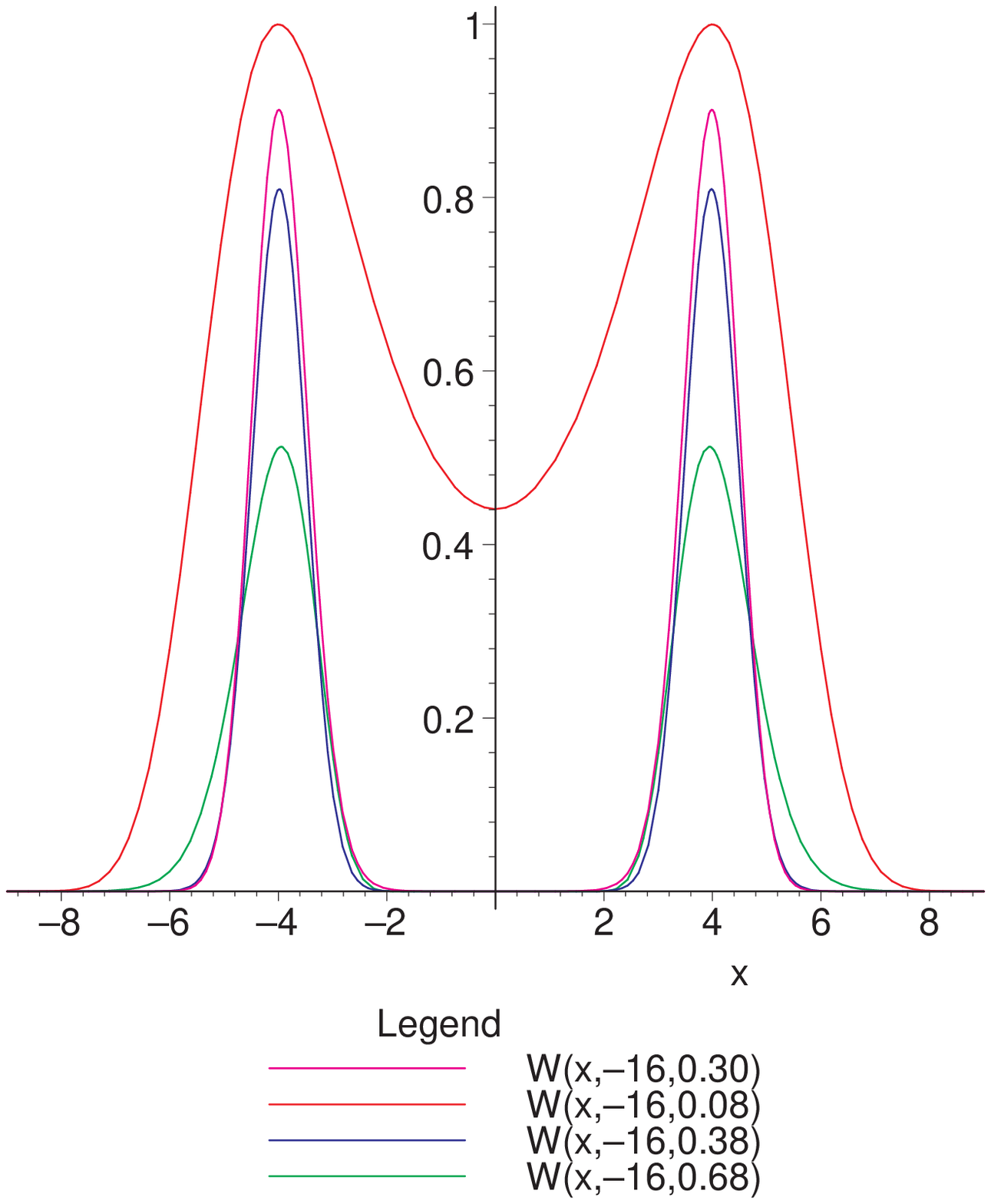}}
\end{center}
\subsection{$\boldsymbol\phi^4$ Theory}
The action is, $S(\phi) = \mu\, \phi^2/2 + g\, \phi^4/4$. Thus, the mollified generating
functional is given by:
\begin{equation*}
  \mathcal{Z}_{\epsilon}[J] = \frac{\displaystyle\varint_{-\infty}^{\infty}\, \ev{\exp\biggl\{i\,
    \frac{\mu}{2}\, \phi^2 + i\, \frac{g}{4}\, \phi^4 - i\, J\, \phi\biggr\}}_{\epsilon}\,
    d\phi}{\displaystyle\varint_{-\infty}^{\infty}\, \ev{\exp\biggl\{i\, \frac{\mu}{2}\, \phi^2
    + i\, \frac{g}{4}\, \phi^4 \biggr\}}_{\epsilon}\, d\phi} \; .
\end{equation*}
As usual, its parameters, $(m,g)$, will determine the phase structure of the theory. The
following are the plots of the 2-point Green's function in the symmetric and
broken-symmetric phases, respectively.
\begin{center}
  \hspace{\fill}
  \scalebox{0.2}{\includegraphics{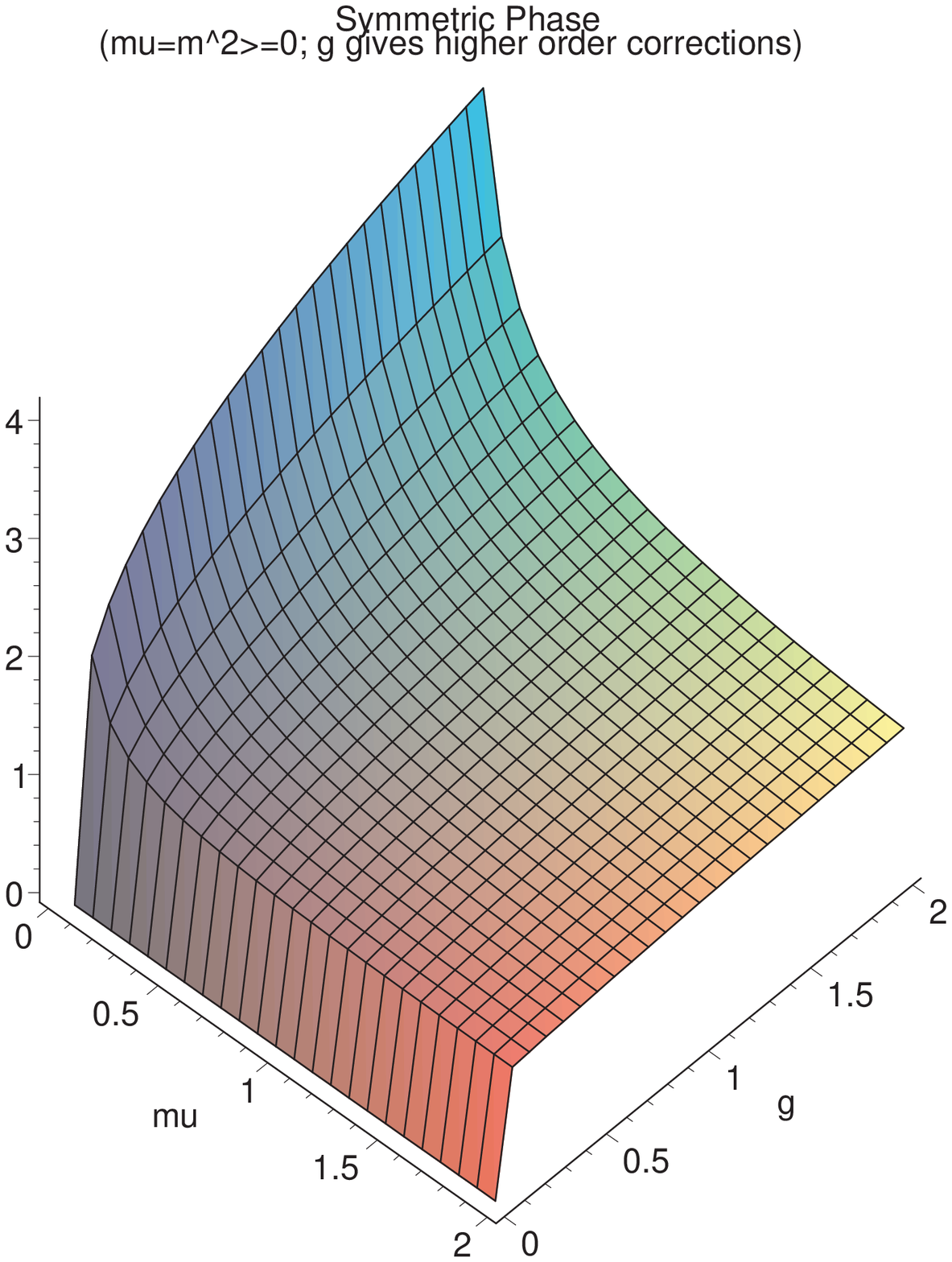}}
  \hspace{\fill}
  \scalebox{0.2}{\includegraphics{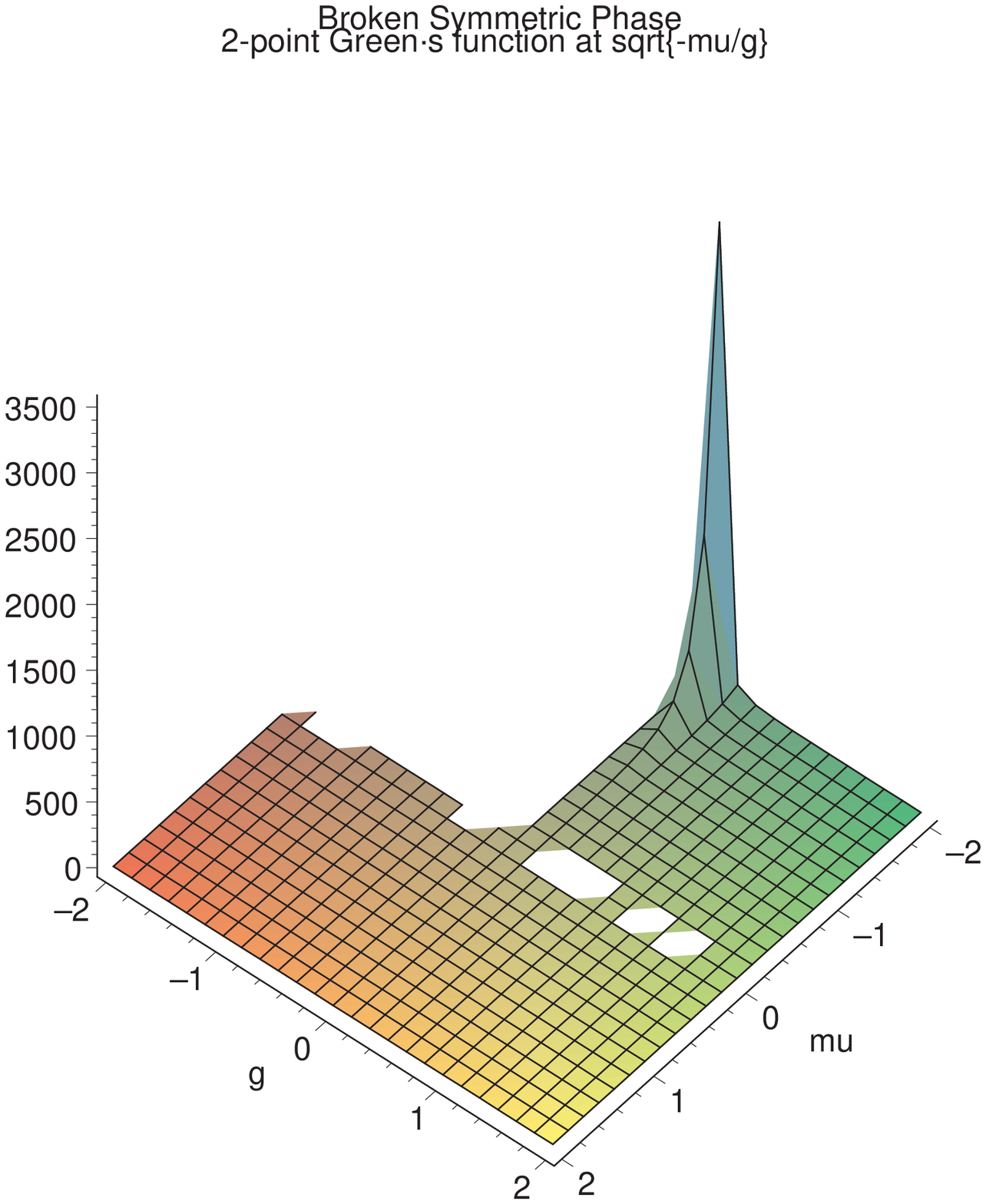}}
  \hspace{\fill}
\end{center}
The plots below are the importance function and the mollified integrand in the symmetric
and broken symmetric phase, respectively.
\begin{center}
  \hspace{\fill}
  \scalebox{0.2}{\includegraphics{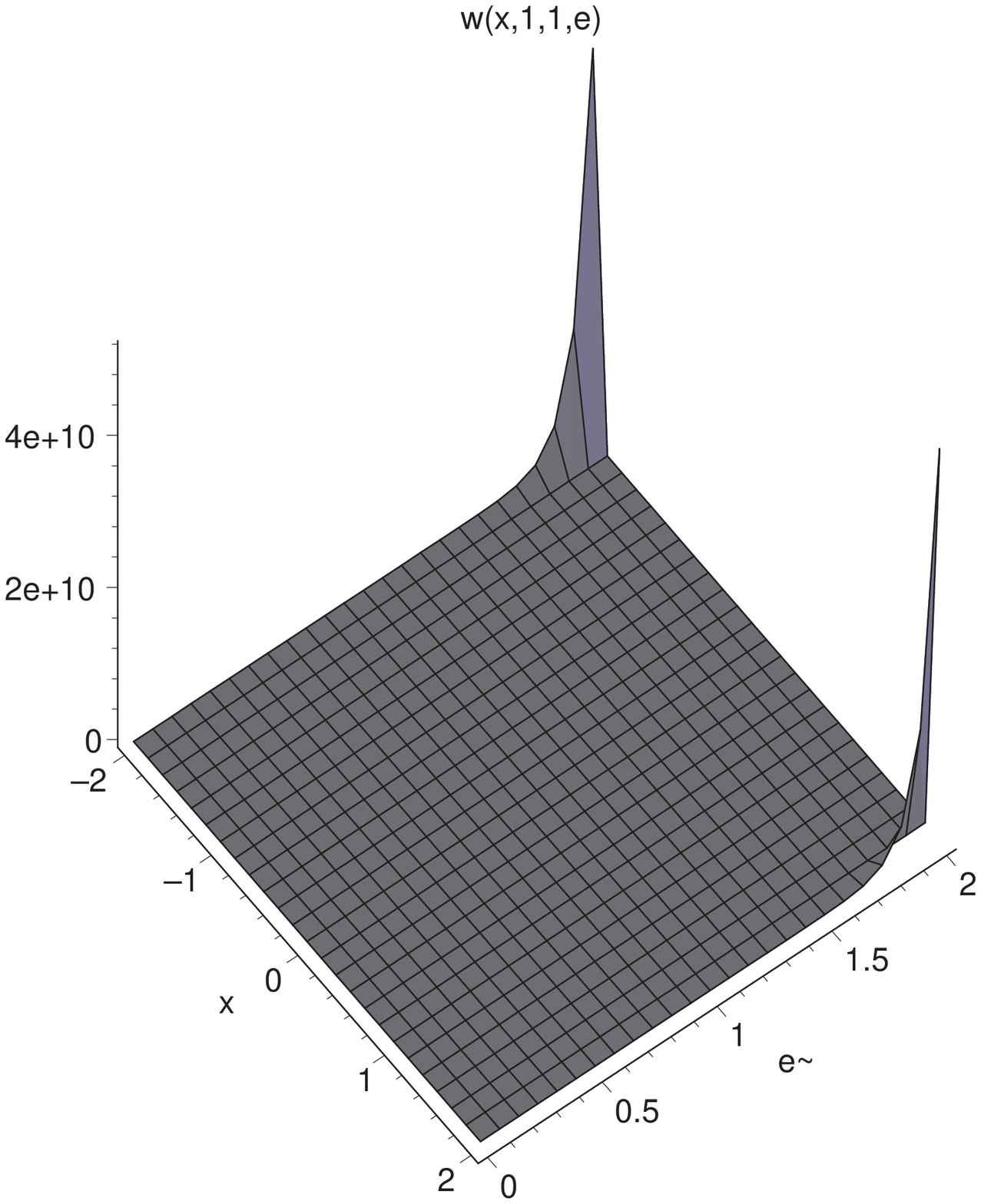}}
  \hspace{\fill}
  \scalebox{0.2}{\includegraphics{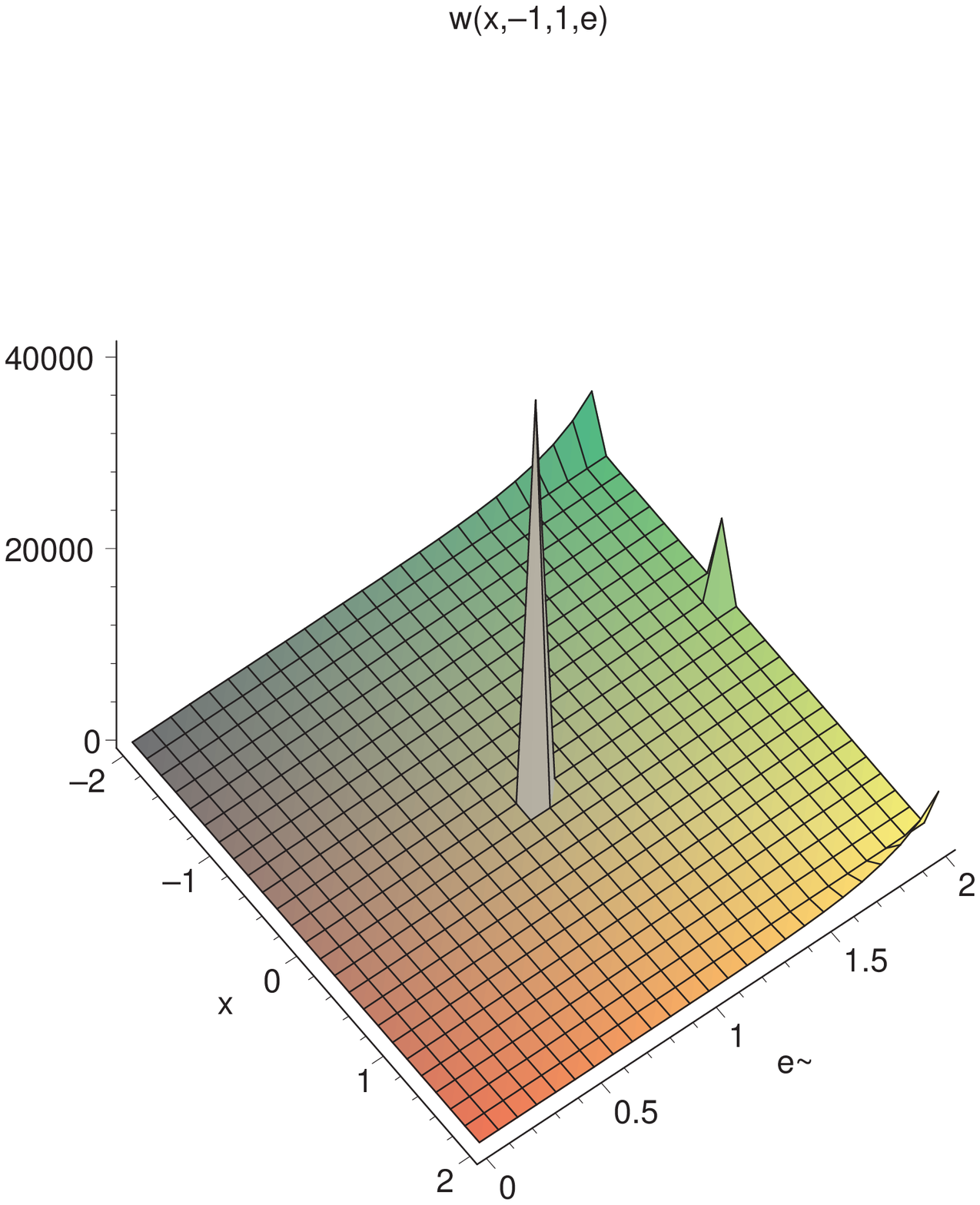}}
  \hspace{\fill}
  \hspace{\fill}
  \scalebox{0.2}{\includegraphics{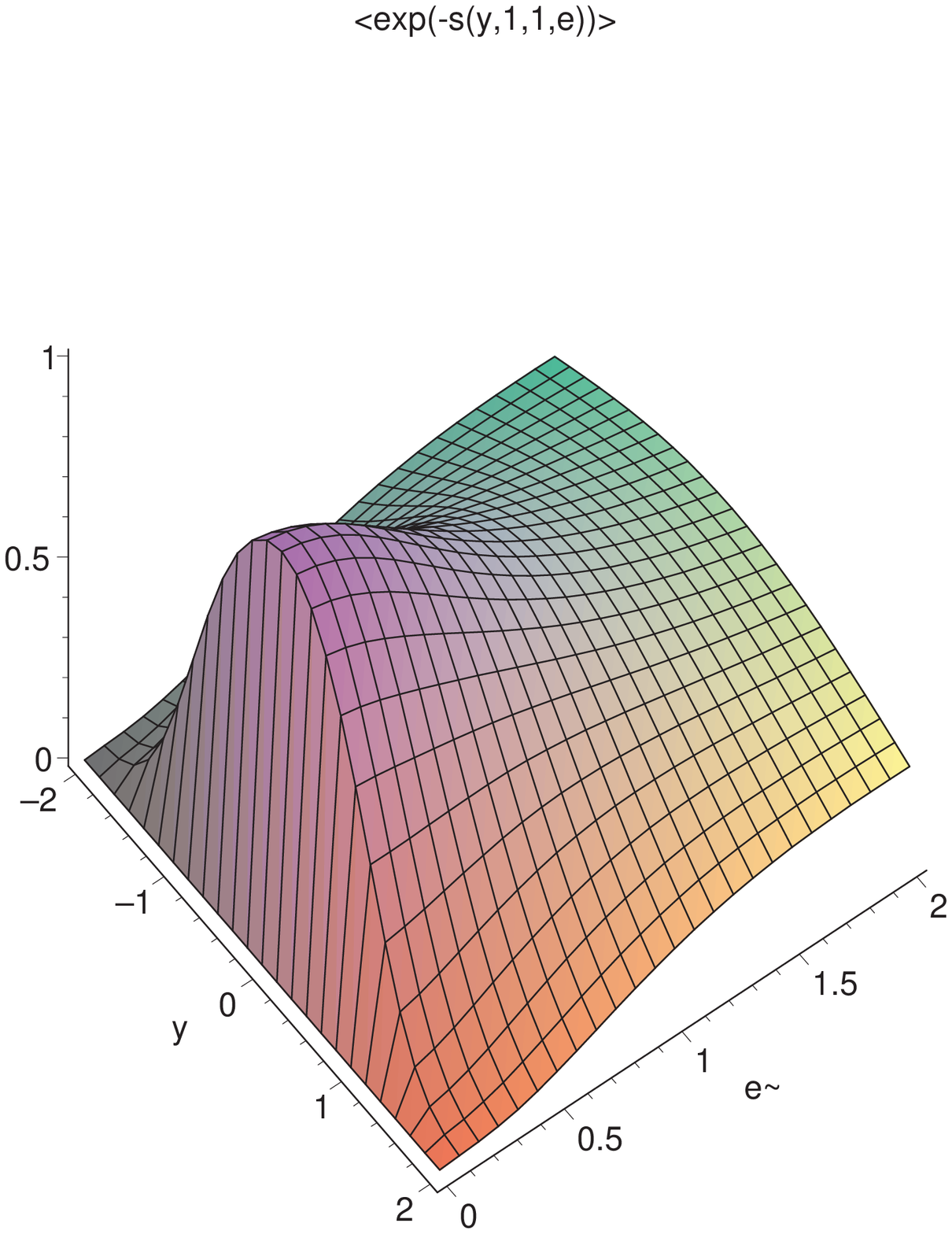}}
  \hspace{\fill}
  \scalebox{0.2}{\includegraphics{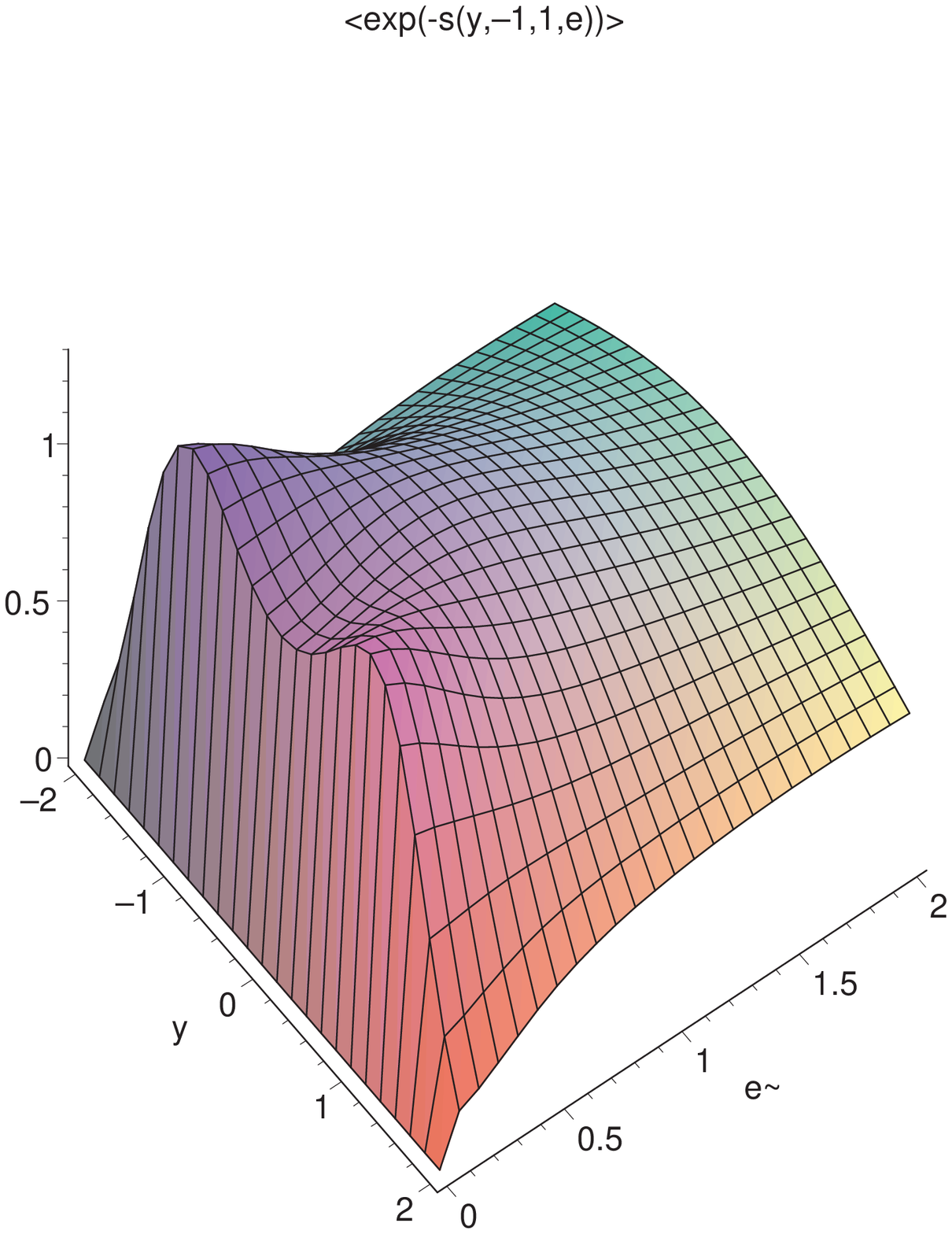}}
  \hspace{\fill}
\end{center}
It is interesting to realize that, for the broken symmetric phase, the importance function
peaks exactly at the point where the mollified integrand ceases to have two concavities
and begins to have only one.
\section{Acknowledgements}
 DDF and GSG wish to acknowledge support by \textsf{DOE} grant \textsf{DE-FG02-91ER40688 - Task
  D}\/, JDD and DS wish to acknowledge support from the \textsf{National Science Foundation}
  through awards \textsf{CHE-0095053} and \textsf{CHE-0131114}\/.
\vspace{1pc}

\begin{thebibliography}{11}
  \bibitem{pde} L. C. Evans, \emph{Partial Differential Equations}, AMS (1998), GSM 19.
  \bibitem{fismath} M. Reed and B. Simon, \emph{Methods of Modern Mathematical Physics II:
      Fourier Analysis, Self-Adjointness}, Academic Press.
 \bibitem{doll} D. Sabo and J.D. Doll and D.L. Freeman, \emph{Stationary Tempering and the
     Complex Quadrature Problem}, J. Chem. Phys., 116(9):3509-3520, March 2002.
 \bibitem{gerry} S. Garcia and Z. Guralnik and G. Guralnik, \emph{Theta Vacua and Boundary
     Conditions of the Schwinger-Dyson Equations}, (1996), \texttt{hep-th/9612079}.
\end{thebibliography}
\end{document}